\documentclass[twocolumn,showpacs,preprintnumbers,amsmath,amssymb]{revtex4}

\def\undersim#1{\setbox9\hbox{${#1}$}{#1}\kern-\wd9\lower
    2.5pt \hbox{\lower\dp9\hbox to \wd9{\hss $_\sim$\hss}}}
\usepackage{amssymb}
\usepackage{amsmath}
\usepackage{graphicx}
\usepackage{color}

\def\undersim#1{\setbox9\hbox{${#1}$}{#1}\kern-\wd9\lower
    2.5pt \hbox{\lower\dp9\hbox to \wd9{\hss $_\sim$\hss}}}

\def\mk{{\mathbf p}}

\begin{document}

\title{Squeezed back-to-back correlations of bosons with nonzero widths in
relativistic heavy-ion collisions}

\author{Peng-Zhi Xu$^1$}
\author{Wei-Ning Zhang$^{1,2}$\footnote{wnzhang@hit.edu.cn, wnzhang@dlut.edu.cn}}
\affiliation{$^1$Department of Physics, Harbin Institute of Technology, Harbin,
Heilongjiang 150006, China\\
$^2$School of Physics, Dalian University of Technology, Dalian, Liaoning 116024,
China}


\begin{abstract}
We derive the formulas for calculating the squeezed back-to-back correlation
(SBBC) between a boson and antiboson with nonzero width produced in relativistic
heavy-ion collisions. The SBBCs of $D^0$ and $\phi$ mesons with finite in-medium
widths are studied. We find that the finite width can change the pattern of the
SBBC function of $D^0{\bar D}^0$ with respect to mass. However, the SBBC function
of $\phi\phi$ is insensitive to the width. In the high-particle-momentum region,
the SBBC function of $\phi\phi$ increases with particle momentum rapidly and
can exceed that of $D^0{\bar D}^0$ whether the width is nonzero or not.
\end{abstract}
\maketitle

\section{Introduction}
The interaction of particles with a medium in relativistic heavy-ion collisions may
cause a squeezed back-to-back correlation (SBBC) of detected boson-antiboson pairs
\cite{AsaCso96,AsaCsoGyu99,Padula06,DudPad10,Zhang15a,Zhang-EPJC16,AGY17,XuZhang19}.
This SBBC is related to the in-medium mass modification of the bosons through
a Bogoliubov transformation between the annihilation (creation) operators of
the quasiparticles in the medium and the corresponding free particles
\cite{AsaCso96,AsaCsoGyu99,Padula06,DudPad10,Zhang15a,Zhang-EPJC16,AGY17,XuZhang19}.
Generally, the in-medium mass modification includes not only a mass shift of the
boson but also an increase of its width in the medium. Therefore, the necessity
of developing a formulism that can be used to calculate the SBBC between a boson
and antiboson with nonzero width in a medium is obvious.

In this work, we derive the formulas for calculating the SBBC function between
a boson and antiboson with nonzero width. The influences of the in-medium width
on the SBBC functions of $D^0{\bar D}^0$ and $\phi\phi$ are investigated. We find
that the SBBC function of $D^0{\bar D}^0$ changes obviously for a finite change
of width. The SBBC of $\phi\phi$ increases with increasing particle momentum
rapidly in a high-momentum region and can exceed the SBBC of $D^0{\bar D}^0$ at
high momenta whether the width is nonzero or not. Because of the presence of a
charm or strange quark, which is believed to experience the entire evolution of
the quark-gluon plasma (QGP) created in relativistic heavy-ion collisions, the
analyses of experimental data of $D$ and $\phi$ mesons have recently attracted
great interest \cite{{STAR-PRC19D,CMS-PRL18D,CMS-PLB18D,ALICE-JHEP18D,
ALICE-JHEP16D,ALICE-JHEP15D,ALICE-PRC14D,ALICE-PRL13D,ALICE-EPJC18p,STAR-PRL16p,
STAR-PRC16p,PHENIX-PRC16p,ALICE-PRC15p,PHENIX-PRC11p,STAR-PLB09p,STAR-PRC09p,
STAR-PRL07p,PHENIX-PRL07p,PHENIX-PRC05p,NA50-PLB03p,STAR-PRC02p,NA50-PLB00p}}.
However, the bosons with large masses have strong SBBC
\cite{Padula-JPG10,Zhang-IJMPE15,Zhang-CPC15,Zhang-EPJC16}. The study of the
heavy-meson SBBC is meaningful in relativistic heavy-ion collisions.

The rest of this paper is organized as follows. In Section II, we present the
formula derivations of the SBBC function for a boson and antiboson with nonzero
in-medium width. Then, we show the results of the SBBC functions of $D^0{\bar D}^0$
and $\phi\phi$ in Section III.  Finally, a summary is given in Section IV.

\section{Formulas}
For a system of a boson with mass $m_0$ in vacuum, the Hamiltonian density is
given by
\begin{equation}
{\cal H}_0(x)=\frac{1}{2}\left\{{\dot\phi}^2(x)+[\nabla\phi(x)]^2 +m_0^2
\phi^2(x)\right\},
\end{equation}
where
\begin{equation}
\phi(x)=\sum_{\mk}(2V\omega_\mk)^{-\frac{1}{2}}\left(e^{-ip\cdot x}a_\mk
+e^{ip\cdot x}a^\dag_\mk \, \right),
\end{equation}
where $a_\mk$ and $a^\dag_\mk$ are annihilation and creation operators
of the free boson, respectively, $p=(\omega_\mk,\mk)$, and $\omega_\mk=
\sqrt{\mk^2+m_0^2}$.

Denoting the in-medium mass shift and width as $\Delta m$ and $\Gamma$,
respectively, the boson in-medium energy can be written as
\begin{equation}
\Omega_\mk =\sqrt{\mk^2+(m_0+\Delta m -i\Gamma/2)^2}\equiv |\Omega_\mk|\,
e^{i\Theta},
\end{equation}
where
\begin{equation}
|\Omega_\mk|=\left\{\bigg[\mk^2\!+\!(m_0+\Delta m)^2\!-\!\frac{\Gamma^2}{4}
\bigg]^2\!+\!(m_0\!+\!\Delta m)^2\Gamma^2\right\}^{\!{1/4}},
\end{equation}
\begin{equation}
\Theta=\frac{1}{2}\tan^{-1}\left[\frac{-(m_0+\Delta m)\Gamma}
{\mk^2\!+\!(m_0+\Delta m)^2\!-\!\Gamma^2/4}\right].
\end{equation}
Here, $\Theta<0$, indicating the imaginary part of $\Omega_\mk$ is negative.
The in-medium system Hamiltonian is given by \cite{AsaCso96}
\begin{eqnarray}
&&H_{\rm M}=\!\int \!d^3x\,\frac{1}{2}\left\{{\dot\phi}^2(x)+[\nabla\phi(x)]^2
+(m_0^2+m_1^2)\phi^2(x)\right\}\nonumber\\
&&\hspace*{6mm}=\sum_\mk \omega_\mk a_\mk^{\dag}a_\mk +\frac{1}{4}\sum_\mk
\frac{m_1^2}{\omega_\mk}\bigg[e^{-i2\omega_\mk t}a_\mk a_{-\mk}~~~~~~~~\nonumber\\
&&\hspace*{10mm}+\,e^{i2\omega_\mk t}a_\mk^{\dag}a_{-\mk}^{\dag}+2a_\mk^{\dag}a_\mk
\bigg],
\label{HM1}
\end{eqnarray}
where
\begin{equation}
m_1^2=(m_0+\Delta m -i\Gamma/2)^2-m_0^2=\Omega_\mk^2-\omega_\mk^2.
\end{equation}
Eq.~(\ref{HM1}) reduces to the case in Ref. \cite{AsaCso96} when $\Gamma=0$.

To diagonalize $H_{\rm M}$, we perform the transformation
\begin{equation}
e^{-i\omega_\mk t}a_\mk=c_\mk e^{-i\Omega_\mk t}b_\mk +s^*_{-\mk} e^{i\Omega_\mk
t}b^{\dag}_{-\mk},
\label{B-trans}
\end{equation}
\begin{equation}
e^{i\omega_\mk t}a^{\dag}_\mk=c^*_\mk e^{i\Omega_\mk t}b^{\dag}_\mk +s_{-\mk}
e^{-i\Omega_\mk t}b_{-\mk},
\label{B-trans1}
\end{equation}
and obtain
\begin{eqnarray}
&&H_{\rm M}=\frac{1}{2}\sum_\mk \frac{m_1^2}{\omega_\mk}\bigg[(|c_\mk|^2
+|s_\mk|^2+c_\mk s^*_\mk +c^*_\mk s_\mk)~~~~~~\nonumber\\
&&\hspace*{10mm}+\frac{2\omega_\mk^2}{m_1^2}(|c_\mk|^2+|s_\mk|^2)
\bigg]b^\dag_\mk b_\mk \nonumber\\
&&\hspace*{8mm}+\frac{1}{4}\sum_\mk \frac{m_1^2}{\omega_\mk}\bigg[(c^*_\mk
c^*_{-\mk} +s^*_\mk s^*_{-\mk} +2c^*_\mk s^*_{-\mk})~~~~~~\nonumber\\
&&\hspace*{10mm}+\frac{4\omega_\mk^2}{m_1^2}\,c^*_\mk s^*_{-\mk}\bigg]
e^{i2\Omega_\mk t} b^\dag_\mk b^\dag_{-\mk}.
\label{Hm1}
\end{eqnarray}
Here, the time-decayed term of $e^{-i2\Omega_\mk t}b_\mk b_{-\mk}$ has been
removed. In Eqs.~(\ref{B-trans}) and (\ref{B-trans1}), the transformation
coefficients satisfy
\begin{equation}
|c_\mk|^2-|s_\mk|^2=1
\label{cs1}
\end{equation}
to keep the operators $b$ and $b^\dag$ with the same commutation relation
as that of $a$ and $a^\dag$.

Letting the terms $b^\dag_\mk b^\dag_{-\mk}$ and $b^\dag_\mk b_\mk$ in
Eq.~(\ref{Hm1}) equal 0 and $\Omega_\mk b^\dag_\mk b_\mk$, respectively, we
have
\begin{equation}
c_\mk=\frac{\cosh r_1 +i\cosh r_2}{\sqrt{2}},~~
s_\mk=\frac{\sinh r_1 +i\sinh r_2}{\sqrt{2}},
\label{cpsp}
\end{equation}
where $c_\mk=c_{-\mk}$, $s_\mk=s_{-\mk}$, $r_1$ and $r_2$ are two real
functions,
\begin{equation}
r_{1,2}=\frac{1}{2}\ln\left[\frac{\omega_\mk(1\mp \sin\Theta)}
{|\Omega_\mk|\cos\Theta}\right],
\label{r12}
\end{equation}
and therefore
\begin{equation}
H_{\rm M}=\sum_\mk \Omega_\mk b_\mk^{\dag} b_\mk,
\end{equation}
i.e., $b_\mk$ and $b^\dag_\mk$ are annihilation and creation operators,
respectively, of the quasiparticle in the medium with energy $\Omega_\mk$.
For $\Gamma=0$, we have $r_1=r_2=\frac{1}{2}\ln[\omega_\mk/\sqrt{\mk^2+
(m_0+\Delta m)^2}]$, and the diagonalization issue reduces to that for the
zero-width case.

Next, we further consider the boson with a width $\Gamma_0$ in vacuum.
In this case, the boson energy in vacuum is
\begin{equation}
\omega'_\mk=\sqrt{\mk^2+(m_0-i\Gamma_0/2)^2}\equiv |\omega'_\mk| e^{i\theta}.
\end{equation}
Introducing the annihilation and creation operators, $a'_\mk$ and $a'^\dag_\mk$,
respectively, by a transformation similar to Eqs.~(\ref{B-trans}) and
(\ref{B-trans1}) ($a'_\mk \to b_\mk,\,a'^\dag_\mk \to b^\dag_\mk,\omega'_\mk
\to \Omega_\mk$), we can diagonalize the Hamiltonian of the boson with $\Gamma_0$,
and therefore write the Hamiltonian density in this case as
\begin{equation}
{\cal H}'_0(x)=\frac{1}{2}\left\{{\dot\phi}'^2(x)+[\nabla\phi'(x)]^2
+(m_0-i\Gamma_0/2)^2\phi'^2(x)\right\},
\end{equation}
where
\begin{equation}
\phi'(x)=\sum_{\mk}(2V\omega'_\mk)^{-\frac{1}{2}}\left(e^{-ip'\cdot x}a'_\mk
+e^{ip'\cdot x}a'^\dag_\mk \, \right),
\end{equation}
where $p'=(\omega'_\mk,\mk)$.

Again, with the transformation similar to Eqs.~(\ref{B-trans}) and (\ref{B-trans1}),
\begin{equation}
e^{-i\omega'_\mk t}a'_\mk=c_\mk e^{-i\Omega_\mk t}b'_\mk +s^*_{-\mk}
e^{i\Omega_\mk t} b'^{\dag}_{-\mk},
\label{B-transp}
\end{equation}
\begin{equation}
e^{i\omega'_\mk t}a'^{\dag}_\mk=c^*_\mk e^{i\Omega_\mk t}b'^{\dag}_\mk
+s_{-\mk} e^{-i\Omega_\mk t}b'_{-\mk},
\label{B-transp1}
\end{equation}
we can diagonalize the in-medium Hamiltonian
\begin{eqnarray}
&&H'_{\rm M}=\sum_\mk \omega'_\mk a'^\dag_\mk a'_\mk +\frac{1}{4}\sum_\mk
\frac{m'^2_1}{\omega'_\mk}\bigg[e^{-i2\omega'_\mk t}a'_\mk a'_{-\mk}~~~~~~~~~~
\nonumber\\
&&\hspace*{2mm}+\,e^{i2\omega'_\mk t}a'^\dag_\mk a'^\dag_{-\mk}+2a'^\dag_\mk
a'_\mk \bigg],~~~(m'^2_1=\Omega^2_\mk-\omega'^2_\mk)
\end{eqnarray}
to
\begin{equation}
H'_{\rm M}=\sum_\mk \Omega_\mk b'^\dag_\mk b'_\mk.
\end{equation}
In Eqs.~(\ref{B-transp}) and (\ref{B-transp1}), the transformation coefficients
$c_{\mk}$ and $s_{\mk}$ have the same expressions as Eq.~(\ref{cpsp}), but $r_1$
and $r_2$ are now
\begin{equation}
r_{1,2}=\frac{1}{2}\ln\left[\frac{|\omega'_\mk|(1\mp \sin(\Theta-\theta))}
{|\Omega_\mk|\cos(\Theta-\theta)}\right].
\label{r12n}
\end{equation}

The SBBC function of the boson-antiboson with momenta $\mk_1$ and $\mk_2$ is
defined as 
\cite{AsaCsoGyu99,Padula06,DudPad10,Zhang15a,Zhang-EPJC16,AGY17,XuZhang19}
\begin{equation}
\label{BBCf}
C(\mk_1,\mk_2) = 1 + \frac{|G_s(\mk_1,\mk_2)|^2}{G_c(\mk_1,\mk_1) G_c(\mk_2,
\mk_2)},
\end{equation}
where $G_c(\mk_1,\mk_2)$ and $G_s(\mk_1,\mk_2)$ are the chaotic and squeezed
amplitudes, respectively, and
\begin{equation}
G_c(\mk_1,\mk_2)=\sqrt{\omega_{\mk_1}\omega_{\mk_2}}\,\langle a^{\dagger}_{\mk_1}
a_{\mk_2}\rangle ,
\end{equation}
\begin{equation}
G_s(\mk_1,\mk_2)=\sqrt{\omega_{\mk_1}\omega_{\mk_2}}\,\langle a_{\mk_1}
a_{\mk_2}\rangle,
\end{equation}
where $\langle\cdots\rangle$ indicates the ensemble average and $\omega_\mk$
is the energy of a boson with average mass $m_0$ for nonzero $\Gamma_0$.
The SBBC function for a spatially homogeneous source can be written as
\cite{{AsaCsoGyu99,Padula06,Zhang15a,XuZhang19,Zhang-EPJC16,AGY17}}

\begin{figure*}[thbp]
\includegraphics[scale=0.55]{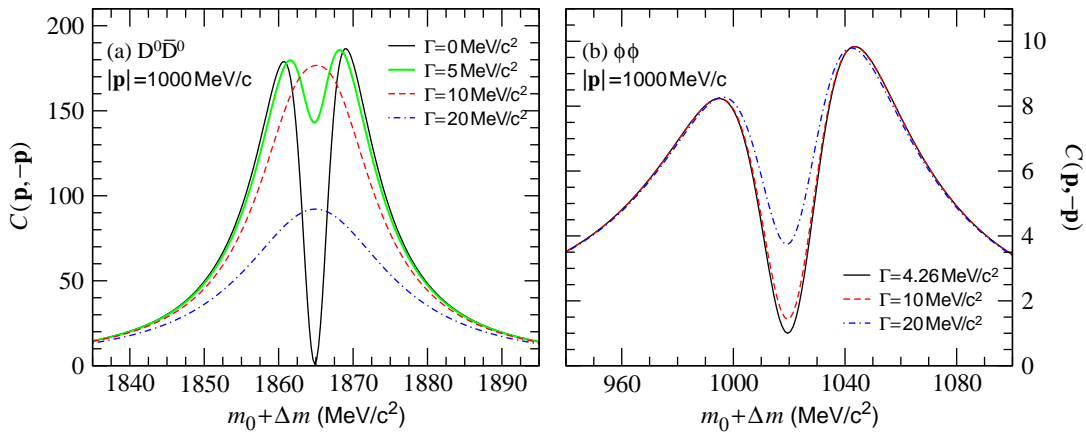}
\caption{SBBC functions of $D^0\!{\bar D}^0$ [(a)] and
$\phi\phi$ [(b)] with respect to in-medium particle mass $(m_0+\Delta m)$
for particle momentum $|\mk|=1000$~MeV/$c$ and different $\Gamma$ values. }
\label{zfSBBCm}
\end{figure*}

\vspace*{-5mm}
\begin{eqnarray}
\label{BBCf1}
C(\mk,-\mk)\!=\!1\!+\!\frac{|c_\mk s^*_\mk n_\mk\!+\!c_{-\mk} s_{-\mk}^*
(n_{-\mk}\!+\!1)|^2}{n_1(\mk)n_1(-\mk)}|{\widetilde F}(\omega_\mk,\Delta t)|^2,
\hspace*{-5mm}\cr
\end{eqnarray}
where $n_\mk$ is the Bose-Einstein distribution of the quasiparticle with the
energy corresponding to in-medium average mass ($m_0\!+\!\Delta m$), $n_1(\mk)\!
=\!|c_{\mk}|^2 n_{\mk}\!+\!|s_{-\mk}|^2(n_{-\mk}\!+\!1)$, and $|{\widetilde F}(
\omega_\mk,\Delta t)|^2$ is a time suppression factor. We take $|{\widetilde 
F}(\omega_\mk,\Delta t)|^2\!=\!(1+4\omega_{\mk}^2 \Delta t^2)^{-1}$ in the 
calculations for a time-profile function of exponential decay as in Refs.
\cite{AsaCsoGyu99,Padula06,DudPad10,Zhang-CPC15,Zhang-IJMPE15,XuZhang19}. 
Generally, $|{\widetilde F}(\omega_\mk,\Delta t)|^2$ is also related to the 
spatial distribution of particle-emitting source for an evolving system
\cite{Zhang15a,Zhang-EPJC16}, and there is a large difference between the 
suppression factors from different time-profile functions 
\cite{Zhang15a,Knoll-11}.

\section{Results}
We plot in Figs.~\ref{zfSBBCm}(a) and \ref{zfSBBCm}(b) the SBBC functions of
$D^0\!{\bar D}^0$ and $\phi\phi$ with respect to in-medium mass $m_0+\Delta m$
for different $\Gamma$ values, respectively. Here, the particle momentum is
fixed at 1000~MeV/$c$ and we take $\Delta t=2$~fm/$c$ in calculations as in
Refs. \cite{AsaCsoGyu99,Padula06,DudPad10,Zhang15a,XuZhang19}. The mass and
width of $D^0$ meson in vacuum, $m_0$ and $\Gamma_0$, are taken to be 1864.86
and 0 MeV/$c^2$ (PDG: $1.60\times 10^{-9}$~MeV/$c$, corresponding to a mean 
life $4.10 \times 10^{-15}$~s) respectively, and the
$m_0$ and $\Gamma_0$ of $\phi$ meson are taken to be 1019.46 and 4.26
MeV/$c^2$ respectively \cite{PDG-PRD12,PDG-PRD18}.

We see from Fig.~\ref{zfSBBCm}(a) that the pattern of the SBBC function of
$D^0\!{\bar D}^0$ changes significantly with in-medium width $\Gamma$. For
$\Gamma=0$, the SBBC function has a typical two-peak structure
\cite{AsaCsoGyu99,Padula06,DudPad10,XuZhang19}. It is 1 (no correlation)
at $m_0$ ($\Delta m=0$) and approaches 1 when $\Delta m \to \pm \infty$.
However, the two peaks of the SBBC function move to $m_0$ and form one peak
rapidly with increasing $\Gamma$. Then, the peak declines with increasing
$\Gamma$. For $\Gamma\ne0$, the SBBC always exists even though $\Delta m=0$.
By comparing the SBBC functions in Figs.~\ref{zfSBBCm}(a) and \ref{zfSBBCm}(b),
we see that the SBBC functions of $\phi\phi$ with respect to mass are much wider
than those of $D^0\!{\bar D}^0$. This is because the SBBC function becomes wide
with decreasing boson mass \cite{Zhang-EPJC16,Zhang-CPC15,Zhang-IJMPE15}.
We also see that the influence of $\Gamma$ on the SBBC function of $\phi\phi$
is small. Because the SBBC function of $\phi\phi$ has a wide mass distribution,
it is insensitive to a mass-distribution change caused by a change of $\Gamma$.
However, the nonzero $\Gamma_0$ of $\phi$ will also counteract the
effect of $\Gamma$ on the SBBC function [see Eq.~(\ref{r12n}) $\theta\ne0$].

\begin{figure}[htbp]
\includegraphics[scale=0.5]{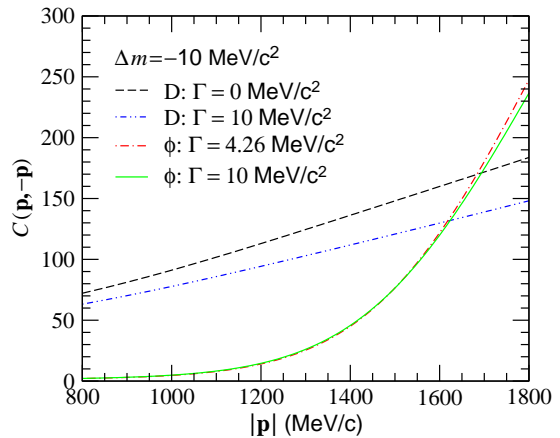}
\caption{SBBC functions of $D^0\!{\bar D}^0$ and $\phi\phi$
with respect to particle momentum for the in-medium mass shift $\Delta
m=-10$~MeV/$c^2$ and in-medium widths $\Gamma=\Gamma_0$ and 10 MeV/$c^2$. }
\label{zfSBBCp}
\end{figure}

We plot in Fig.~\ref{zfSBBCp} the SBBC functions of $D^0\!{\bar D}^0$ and
$\phi\phi$ with respect to particle momentum for the in-medium mass shift
$\Delta m=-10$~MeV/$c^2$ and in medium width $\Gamma=\Gamma_0$ and 10~MeV/$c^2$.
We see that the SBBC functions increase with increasing particle momentum,
and the influence of $\Gamma$ increases with increasing particle momentum.
Because the momentum distribution $n_\mk=n_{-\mk}$ approaches zero when
$|\mk| \to \infty$, the behavior of the SBBC function at very high momenta
is mainly determined by $(|c_\mk s^*_\mk|^2/|s_\mk|^4)$
\cite{AsaCsoGyu99,Zhang-EPJC16}, which is approximately
$16\mk^4/[4m_0^2\Delta m^2+m_0^2(\Gamma-\Gamma_0)^2]$. Therefore, the SBBC
functions of $\phi\phi$ increase with increasing particle momentum more rapidly
than that of $D^0{\bar D}^0$ in the high-momentum region and can exceed the SBBC
of $D^0{\bar D}^0$ at high momenta.

\section{Summary and discussion}
We derived the formulas for calculating the SBBC between a boson and antiboson
with nonzero width produced in relativistic heavy-ion collisions. The influences
of the in-medium width on the SBBC functions of $D^0{\bar D}^0$ and $\phi\phi$
are investigated. It is found that the pattern of the SBBC function of $D^0{\bar
D}^0$ with respect to mass changes significantly with the width. However, the
SBBC function of $\phi\phi$ changes slightly with the width. The influence of
the width on the SBBC increases with particle momentum. Whether the width is
nonzero or not, the SBBC function of $\phi\phi$ increases with increasing particle
momentum more rapidly than that of $D^0{\bar D}^0$ in the high-momentum region and
can exceed the SBBC function of $D^0{\bar D}^0$ at high momenta.

Finally, it is necessary to mention that we have removed the time-decayed
term of $e^{-i2\Omega_\mk t}b_\mk b_{-\mk}$ in diagonalizing the in-medium
Hamiltonian [Eq.~(\ref{Hm1})]. Therefore, the diagonalization is an approximation 
unless the imaginary part of $-2\Omega_\mk$~[\,${\rm Im}(-2\Omega_\mk)\!\sim\!m_0
\Gamma/\omega_\mk\!\sim\!\Gamma$ for~$\mk^2\!<\!m_0^2$\,] is very large.
This problem does not appear in the diagonalization for the bosons without
width. There, the terms of $b_\mk b_{-\mk}$ and $b^\dag_\mk b^\dag_{-\mk}$
can become zero simultaneously with the reduced transform quantity,
$r_1\!\!=\!\!r_2\!\!=\!\!\frac{1}{2}\ln[\,\omega_\mk/\!\sqrt{\mk^2\!
+\!(m_0\!+\!\Delta m)^2}\,]$. The recent measurements of $D^0$ in heavy-ion
collisions at the RHIC and LHC indicate that the average width of $D^0$ is
approximately 30~MeV/$c^2$ \cite{{STAR-PRC19D,CMS-PRL18D,CMS-PLB18D,ALICE-JHEP18D,
ALICE-JHEP16D,ALICE-JHEP15D,ALICE-PRC14D,ALICE-PRL13D}}. The corresponding
characteristic size is $c\tau\sim 6.6$~fm, which is smaller than the typical
size of the particle-emitting source in relativistic heavy-ion collisions.
Therefore, the diagonalization is a good approximation and the influence of 
the in-medium width on the SBBC of $D^0{\bar D}^0$ must be considered in the 
heavy-ion collisions. For the $\phi$ meson, its $c\tau$ is comparable to the 
typical size of the source. Our work is a key step forward to solve the problem. 
In addition, it will be of interest to expand the approach presented in the case 
that the particle and antiparticle with different in-medium mass modifications.

\begin{acknowledgements}
W. N. Zhang thanks Yong Zhang for helpful discussions.
This research was supported by the National Natural Science Foundation of China
under Grant Nos. 11675034 and 11275037.
\end{acknowledgements}

\end{document}